# MODELING OF CABLES WITH HIGH- AND LOW-TENSION ZONES USING A HYBRID ROD-CATENARY FORMULATION

S. Goyal and N.C. Perkins
University of Michigan, Department of Mechanical Engineering
2350 Hayward, Ann Arbor MI-48109-2125 (U.S.A.)
ncp@umich.edu

*Abstract*

*Cables under very low tension may become highly contorted and form loops, tangles, knots and kinks. These nonlinear deformations, which are dominated by flexure and torsion, pose serious concerns for cable deployment. Simulation of the three-dimensional nonlinear dynamics of loop and tangle formation requires a 12th order rod model and the computational effort increases rapidly with increasing cable length and integration time. However, marine cable applications which result in local zones of low-tension very frequently involve large zones of high-tension where the effects of flexure and torsion are insignificant. Simulation of the three-dimensional dynamics of high-tension cables requires only a 6th order catenary model which significantly reduces computational effort relative to a rod model. We propose herein a hybrid computational cable model that employs computationally efficient catenary elements in high-tension zones and rod elements in localized low-tension zones to capture flexure and torsion precisely where needed.*

## 1. INTRODUCTION

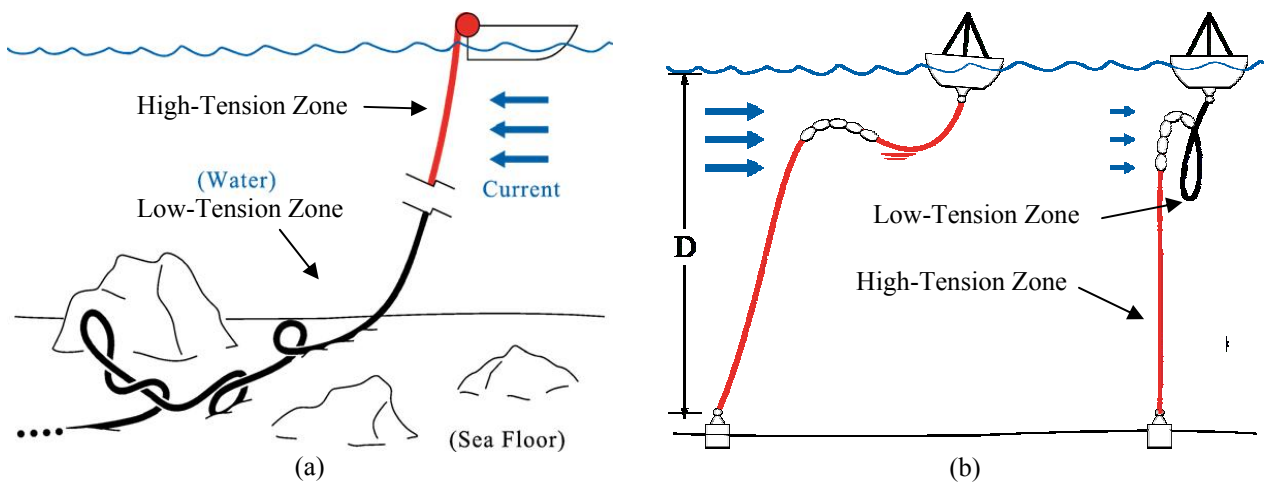

Figure 1: Cables forming loops and tangles in low-tension zones e.g. (a) on the sea floor and (b) in S-tether.

Marine cables tend to form loops and tangles in low tension zones due to combined effects of residual torsion and flexure. This loading scenario is often realized on the seabed or in S-tethers as illustrated in Fig. 1. In this context, loops are often termed "hockles" and these can hinder cable laying and recovery operations, attenuate signal transmission in fiber-optic cables, and can even lead to the formation of knots and kinks that damage cables.

The large nonlinear deformations associated with loop formation are dominated by flexure and torsion, effects that are not captured in models that treat the cable as perfectly flexible. To capture these effects, one must treat the cable as a rod-like element following, for example, the classical rod theory of Kirchhoff/Clebsch [1].

Rod theories have been used to model the mechanics of cables starting with the work of Zajac [2] who studied the onset of "pop-out" instabilities in planar loops using equilibrium rod theory. His work on the nonlinear equilibria of rods has been extended [3-6] to further investigate loop formation and "pop-out" instabilities under a variety of loading scenarios. These instabilities initiate large dynamic responses which may also produce nonlinear transitions to more energetically favorable equilibria. Very few studies [7-12] have developed dynamic rod models sufficient for describing the dynamic evolution of loops. The dynamic model in [12] is artificially damped to study the quasi-static evolution of self-contact and intertwining in biological filaments, while those in [7-11] capture dynamic responses of underwater cables with hydrodynamic forces [13] but without self-contact. This formulation has been extended by Goyal et al. [14-16] to include self-contact and also non-homogeneous and non-isotropic behaviors.

The flexural and torsional effects, however, tend to dominate only in low tension zones of cable. In many underwater cable applications such as depicted in Fig. 1, the low tension zones are confined to a very small fraction of the computational domain rendering the use of a rod model for the entire cable both computationally inefficient and unnecessary. The dynamics of high-tension zones can be efficiently described using a 6th order flexible "catenary". An example of 6th order catenary element was employed by Howell [17] to investigate the dynamics of hanging chains.

Goyal and Perkins [18] proposed an efficient computational cable model that exploits distinct formulations in low- versus high-tension zones. The resulting hybrid cable model adaptively transitions between rod and catenary elements contingent on the relative importance of tension versus bending and/or torsion in any sub-domain. An example of this approach appears in Sun and Leonard [7] who proposed a hybrid computational model of fluid-loaded cables that blends rod and catenary formulations through a point discontinuity. While the model in [7] is designed to solve for only two-domain hybrid cable, the model in [18] can be employed to solve multi-domain hybrid cable. In [7], the spatial integration is done using a shooting method in rod and catenary elements towards the single point discontinuity as depicted by arrows in Fig. 2. In [18], the shooting method is adapted to pass over the point discontinuities to integrate along a single direction. In this paper, we describe the adaptation of shooting method through a specific type of point discontinuity, namely a spherical joint. We open in Section 2 by reviewing the governing equations of the hybrid model [18] with spherical joint constraints.

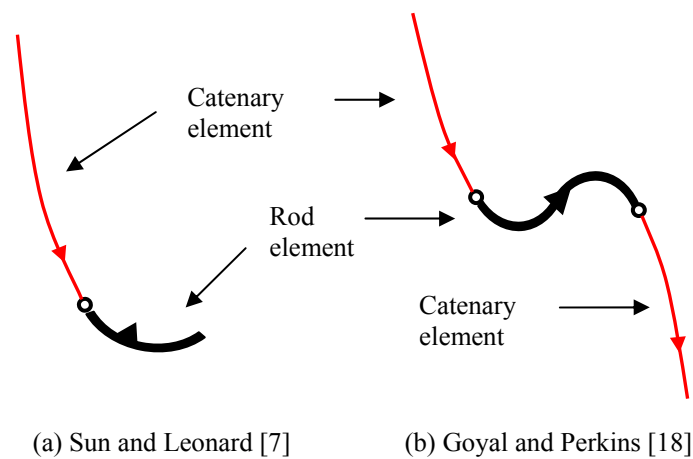

(a) Sun and Leonard [7]   (b) Goyal and Perkins [18]

Figure 2: Integration by shooting method in hybrid models.

## 2. HYBRID ROD-CATENARY MODEL

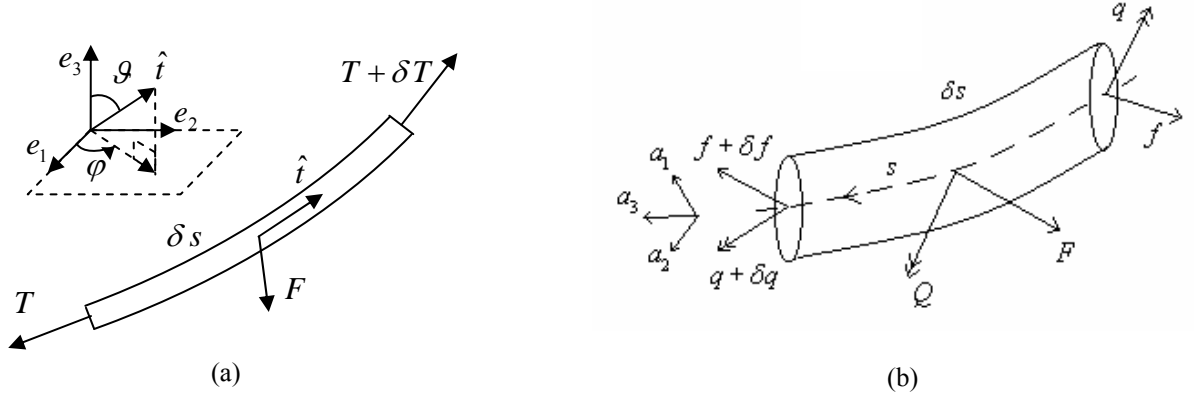

Figure 3: Free body diagrams of an infinitesimal elements of a catenary (a) and rod (b).

The three-dimensional curve formed by the cable centerline is parameterized by the (unstretched) arc-length coordinate $s$ and time $t$ and has the centerline tangent $\hat{t}(s,t)$. The centerline is divided into sub-domains of rod and catenary elements. The infinitesimal elements of rod and catenary are shown in Fig. 3. The rod element employs a body-fixed frame $\{a_i\}$ at each cross-section to describe its orientation with respect to the inertial frame $\{e_i\}$.

### 2.1 Governing Equations for a Catenary

The field equations[1] for the catenary are:

$$\frac{\partial v}{\partial s} = \frac{\partial}{\partial t}\{(1+\varepsilon)\hat{t}\}, \qquad (1)$$

$$\hat{t}\frac{\partial T}{\partial s} + T\frac{\partial \hat{t}}{\partial s} + F = m\frac{\partial v}{\partial t}, \qquad (2)$$

where $m(s)$ denotes the mass per unit arc length, $F(s,t)$ denotes the distributed force per unit arc-length, $v(s,t)$ denotes the velocity of the centerline, and $\varepsilon(s,t)$ denotes the extensional strain related to the tension $T(s,t)$ through the compliance $c_k(s)$,

$$\varepsilon(s,t) = c_k(s)T(s,t). \qquad (3)$$

All derivatives are relative to the inertial frame $\{e_i\}$. The unit vector $\hat{t}$ can be parameterized by its two spherical co-ordinates (see Fig. 3) $\vartheta(s,t)$ and $\varphi(s,t)$ such that its components along $\{e_i\}$ become

$$\hat{t} = [\sin\vartheta\cos\varphi \quad \sin\vartheta\sin\varphi \quad \cos\varphi]. \qquad (4)$$

Substituting Eqs. (3-4) in Eqs. (1-2), yields a 6$^{th}$ order system of partial differential equations in the field variables $Y_c = \begin{Bmatrix} v \\ p \end{Bmatrix}$, where vector $p = \{T \quad \vartheta \quad \varphi\}^T$.

---

[1] Equation (1) results from the continuity of the (extensible) centerline and Eq. (2) results from balance of linear momentum.

## 2.2 Governing Equations for a Rod

The deformation field in rod is represented by the curvature and twist vector $\kappa(s,t)$ that is defined by the rotation per unit arc length of the body-fixed frame. The curvature and twist $\kappa(s,t)$ results in the internal moment $q(s,t)$ in rod element that obeys an (assumed) linear elastic constitutive law:

$$q(s,t) = B(s)\kappa(s,t) \qquad (5)$$

where the tensor $B(s)$ captures the stiffness of the rod in bending and torsion. The rod is assumed to be unshearable and inextensible, but can sustain shear and tensile stresses resulting in the internal force $f(s,t)$.

The field equations[2] for the rod are:

$$\frac{\partial v}{\partial s} + \kappa \times v = \omega \times \hat{t}, \qquad (6)$$

$$\frac{\partial \omega}{\partial s} + \kappa \times \omega = \frac{\partial \kappa}{\partial t}, \qquad (7)$$

$$\frac{\partial q}{\partial s} + \kappa \times q = I\frac{\partial \omega}{\partial t} + \omega \times I\omega + f \times \hat{t} - Q, \qquad (8)$$

$$\frac{\partial f}{\partial s} + \kappa \times f = m\left(\frac{\partial v}{\partial t} + \omega \times v\right) - F, \qquad (9)$$

where $\omega(s,t)$ denotes the cross-section angular velocity, $I(s)$ denotes the tensor of principal mass moments of inertia per unit arc length, $Q(s,t)$ denotes the distributed moment per unit arclength and the remaining variables/ parameters are as in the catenary model. The partial derivatives are all relative to the body-fixed frame $\{a_i\}$. Equations (6-9) result[3] in a 12$^{th}$ order system of partial differential equations in the field variables $Y_r = \{v \quad f \quad \kappa \quad \omega\}^T$.

## 2.3 Constraints at a Rod-Catenary Interface

The rod and catenary elements must now be joined in such a manner that preserves continuity of the displacement (velocity) field and also force and moment equilibrium. These requirements are achieved by introducing a spherical joint at rod-catenary interface. In the formulation above, the catenary variables are represented with components in the inertial frame $\{e_i\}$, while the rod variables are represented with components in the body-fixed frame $\{a_i\}$. Let $L(s,t)$ denote the tensor that transforms a vector from the inertial frame to the body-fixed frame. If the joint is located at $s = s_j$ along the cable centerline, then the above interface conditions become

---
[2] Equation (6) is the inextensibility constraint, Eq. (7) is a compatibility condition, and Eqs. (8-9) are the Newton-Euler equations for an infinitesimal rod element.

[3] Substitute Eq. (5) into Eq. (8). Also recognize that the tangent vector $\hat{t}$ is constant in the body-fixed frame $\{a_i\}$ for an unshearable rod and points along the principal axis for torsion.

$$q_r(s_j,t) = 0 \Rightarrow \kappa_r(s_j,t) = 0, \tag{10}$$

$$v_r(s_j,t) = L_r(s_j,t)v_c(s_j,t), \tag{11}$$

$$f_r(s_j,t) = L_r(s_j,t)\hat{t}_c(s_j,t)T_c(s_j,t), \tag{12}$$

where the subscripts $r$ and $c$ identify rod and catenary variables, respectively.

## 2.4 Numerical Algorithm

For integration of the initial-boundary value problem, we employ the generalized-α method [9, 11, 19] in both space and time to discretize the partial differential equations and solve them using Newton-Raphson method. The resulting implicit algorithm is $2^{nd}$ order accurate, unconditionally stable, and incorporates a (controllable) numerical dissipation parameter. Starting with the initial value $Y(s,0)$, the discretized equations are integrated over space at each successive time step. The boundary conditions are satisfied during spatial integration using a shooting method [7, 10]. The $6^{th}$ order catenary requires 3 boundary conditions at the two boundary points. Similarly, the $12^{th}$ order rod requires 6 boundary conditions at the two boundary points. At the rod-catenary interface, the 9 constraints of the spherical joint given by Eqs. (10-12) substitute for the (6+3) boundary conditions required by each rod and catenary models. In the next section, we demonstrate how the constraints can be imposed to transform the solution space in shooting method through the joint.

## 3. SHOOTING METHOD FOR THE HYBRID MODEL

In shooting method [7, 10], the solution ($Y(s,t)$) is constructed as a linear combination of particular ($Y^P(s,t)$) and homogenous ($Y^H(s,t)$) solutions satisfying the start-point boundary conditions:

$$Y_r(s,t) = \underset{12\times 1}{Y_r^P(s,t)} + \underset{12\times 6}{Y_r^H(s,t)} \underset{6\times 1}{\xi_r(t)}, \tag{13}$$

$$Y_c(s,t) = \underset{6\times 1}{Y_c^P(s,t)} + \underset{6\times 3}{Y_c^H(s,t)} \underset{3\times 1}{\xi_c(t)}, \tag{14}$$

where $\xi$ are arbitrary constants to be determined by matching the end-point boundary conditions. At the joint $s_j$, particular and homogeneous solutions of rod and catenary must fulfill the joint constraints (10-12) linearized about the guessed solution $Y^*(s,t)$. For linearization, we make a simple approximation that $L_r(s_j,t) = L_r^*(s_j,t)$ so that the constraints (10) and (11) are already linear. The constraint (12) upon linearization becomes

$$f_r = L_r^*\left(\hat{t}_c^* T_c^* - \left[\frac{\partial(\hat{t}_c T_c)}{\partial p_c}\right]^* p_c^*\right) + L_r^*\left[\frac{\partial(\hat{t}_c T_c)}{\partial p_c}\right]^* p_c. \tag{15}$$

Next, we partition the matrices of homogeneous and particular solutions as

$$Y_r^H = \begin{bmatrix} {}_1v_r^H & {}_2v_r^H \\ {}_1f_r^H & {}_2f_r^H \\ {}_1\kappa_r^H & {}_2\kappa_r^H \\ {}_1\omega_r^H & {}_2\omega_r^H \end{bmatrix}, \quad Y_r^P = \begin{Bmatrix} v_r^P \\ f_r^P \\ \kappa_r^P \\ \omega_r^P \end{Bmatrix}, \quad Y_c^H = \begin{bmatrix} v_c^H \\ p_c^H \end{bmatrix} \text{ and } Y_c^P = \begin{Bmatrix} v_c^P \\ p_c^P \end{Bmatrix}, \quad (16)$$

where all the elements with superscript $H$ are 3×3 matrices and those with superscript $P$ are 3×1 vectors. We also impose that $\xi_r = \begin{Bmatrix} \xi_c \\ \xi_n \end{Bmatrix}$ where $\xi_n$ is 3×1 vector. Finally, substituting the linear combinations (13,14), the linearized constraints reduce to the following form:

$$\begin{Bmatrix} 0 \\ C \end{Bmatrix} + \begin{bmatrix} A & \Theta \\ \Theta & B \end{bmatrix} \begin{Bmatrix} v_c^P \\ p_c^P \end{Bmatrix} + \begin{bmatrix} A & \Theta \\ \Theta & B \end{bmatrix} \begin{Bmatrix} v_c^H \\ p_c^H \end{Bmatrix} \xi_c = \begin{Bmatrix} v_r^P \\ f_r^P \end{Bmatrix} + \begin{bmatrix} {}_1v_r^H \\ {}_1f_r^H \end{bmatrix} \xi_c + \begin{bmatrix} {}_2v_r^H \\ {}_2f_r^H \end{bmatrix} \xi_n, \quad (17)$$

$$\kappa_r^P + {}_1\kappa_r^H \xi_c + {}_2\kappa_r^H \xi_n = 0, \quad (18)$$

where the 3×3 matrices A and B and the 3×1 vector $C$ are known at the guessed solution and $\Theta$ is null matrix. We impose constraints (17-18) to transform the homogenous and particular solutions of one property element into those of the other at the joint. For example, integrating from catenary domain to rod domain, we need to construct $Y_r^P(s_j,t)$ and $Y_r^H(s_j,t)$ from $Y_c^P(s_j,t)$ and $Y_c^H(s_j,t)$ satisfying Eqs. (17-18) at the joint. We select $\kappa_r^P = {}_1\kappa_r^H = {}_2\kappa_r^H{}_n = {}_2v_r^H = {}_2f_r^H = 0$ so that equating the coefficients of $\xi_n$ in Eq. (17) results in ${}_1v_r^H = Av_c^H$ and ${}_1f_r^H = Bp_c^H$. Equating the remaining terms in Eq. (17) results in $v_r^P = Av_c^P$ and $f_r^P = C + Bp_c^P$. To complete the remaining elements, $\omega_r^P$ can be chosen to be $\omega_r^*$ and ${}_1\omega_r^H$ and ${}_2\omega_r^H$ can be chosen arbitrarily only ensuring that ${}_2\omega_r^H$ is full ranked.

## 4. RESULTS AND DISCUSSION

In this section, we review an example result from [18] featuring a cable suspension with both low and high tension zones. The result validates the hybrid model by benchmarking results with those obtained from a pure rod model. The example is illustrated in Fig. 4 and corresponding parameters are listed in Table 1. The figure shows a cable of length $L$ sagging under its buoyant weight. The left end is pinned. At the right end, the cable centerline is given a prescribed orientation of 30° from the horizontal, is tension-free, but subject to an applied force in the normal (shear) direction. A reaction moment also develops at the right end in response to prescribing the rotation of this end and leads to a region where significant curvature develops.

TABLE 1 - Simulation parameters.

| Quantity | Symbol | Unit | Value | Quantity | Symbol | Unit | Value |
|---|---|---|---|---|---|---|---|
| Diameter | $D$ | m | $1.0 \times 10^{-3}$ | Mass/ length | $m$ | Kg/m | $\pi/4 \times 10^{-3}$ |
| Length | $L$ | m | $2.0 \times 10^{0}$ | Catenary Compliance | $c_k$ | N$^{-1}$ | $0.0 \times 10^{0}$ |
| Bending Moment of Inertia/ length | Tensor $I$ | Kg-m | $\pi/64 \times 10^{-9}$ | Gravitational Acceleration | $g$ | m/s$^2$ | $9.8 \times 10^{0}$ |
| Polar Moment of Inertia/ length | | Kg-m | $\pi/32 \times 10^{-9}$ | Fluid Density | - | Kg/m$^3$ | $1.0 \times 10^{0}$ |
| Bending Stiffness | Tensor $B$ | N-m$^2$ | $5\pi/64 \times 10^{-3}$ | Boundary Force (Fig. 4) | $f_{applied}$ | N | $1.0 \times 10^{-2}$ |
| Torsional Stiffness | | N-m$^2$ | $\pi/16 \times 10^{-3}$ | Joint Location | $s_j$ | m | $1.5 \times 10^{0}$ |

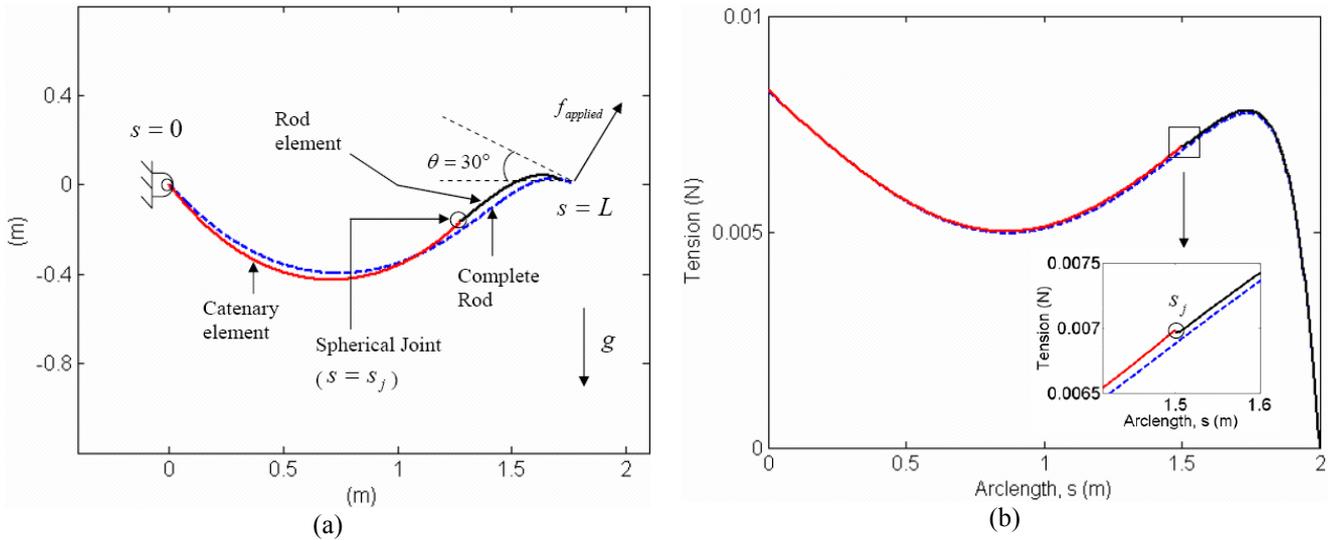

Figure 4: (a) a cable suspension with low (black) and high (red) tension zones. (b) Tension distribution

Two computational models are used to predict the dynamic relaxation to equilibrium from an initially straight cable with hydrodynamic dissipation [13, 18]. The first is the pure rod model (dashed curve) while the second is the hybrid rod-catenary model (solid curve) composed of 25% rod-domain (black) and 75% catenary sub-domain (red). The rod sub-domain is joined to the catenary sub-domain at point $s_j$ with a spherical joint. While the two models predict similar displacements, the modest differences arise from the added flexibility of the hybrid model as seen, for instance by comparing the maximum sag. More importantly, there is a significant increase in the calculation speed for the hybrid model in proportion to the ratio of the catenary/rod sub-domains. For example, under equal conditions, the computational speed is increased by approximately a factor of 1.5 to 2 for this very simple example when using the hybrid model. In addition, the hybrid model successfully captures the mechanics of bending at the right end that would otherwise be missed in a pure catenary model.

## 5. SUMMARY AND CONCLUSIONS

This paper describes a hybrid rod-catenary model that can efficiently compute the dynamics of long cables with multiple low and high tension subdomains. Such tension changes arise in ocean engineering cables, as seen for example, in S-tether moorings and in cases of cable/seabed contact. Pure rod models ultimately lead to ill-conditioned computations for the very flexible, long cables often used in ocean engineering applications. Pure catenary models are well-known to be ill-conditioned for low tension cables. Thus, it is natural to employ these models in opposite tension regimes; that is, employ the rod model in low tension subdomains, and employ the catenary model in high tension subdomains. The hybrid model herein inherits the computational benefits of each model and offers tremendous computational advantages. For the simple cable suspension chosen as an example herein, the hybrid model increases computational speed by a factor of 1.5-2 relative to the pure rod model. In addition, the hybrid model successfully captures the dominant effects of bending in the low tension region that would otherwise be lost in using a pure catenary model.

## ACKNOWLEDGMENTS

The authors gratefully acknowledge the research support provided by the U.S. Office of Naval Research.